\RequirePackage{fix-cm}

\documentclass[natbib, smallextended, a4paper]{svjour3}       

\usepackage{graphicx}
\usepackage{url}
\usepackage{enumitem}
\usepackage{caption}
\usepackage{float}
\usepackage[a4paper, margin=1.25in]{geometry}

\makeatletter
\def\makeheadbox{\relax}
\makeatother

\begin{document}

\title{Visual and Textual Programming Languages: A Systematic Review of the Literature \thanks{This work was undertaken with help from funding by the John and Pat Hume Scholarship, Maynooth University.}}

\author{Mark Noone \and Aidan Mooney \textsuperscript{S,C}}

\institute{Mark Noone \at
	Department of Computer Science, Maynooth University, Maynooth, Co. Kildare, Ireland \\
	\email{mark.noone@mu.ie} \\
	\and
	Dr. Aidan Mooney \at
	Department of Computer Science, Maynooth University, Maynooth, Co. Kildare, Ireland \\
	\email{aidan.mooney@mu.ie \\ S = Senior Author, C = Corresponding Author}
}

\date{2nd December 2017}

\maketitle

\begin{abstract}
	
	It is well documented, and has been the topic of much research, that Computer Science courses tend to have higher than average drop out rates at third level, particularly so for students advancing from first year to second year. This is a problem that needs to be addressed with urgency but also caution. The required number of Computer Science graduates is growing every year but the number of graduates is not meeting this demand and one way that this problem can be alleviated is to encourage students, at an early age, towards studying Computer Science courses. \\
	
	This paper presents a systematic literature review that examines the role of visual and textual programming languages when learning to program, particularly as a first programming language. The approach is systematic, in that a structured search of electronic resources has been conducted, and the results are presented and quantitatively analysed. This study will provide insight into whether or not the current approaches to teaching young learners programming are viable, and examines what we can do to increase the interest and retention of these students as they progress through their education. \\
	
\end{abstract}

\newpage

\section{Introduction and Motivation}
The usage of Computer Science knowledge is becoming much more prevalent in society today. In Ireland, a high number of technology companies choose to set up due to the quality of our third-level Computer Science graduates. However, the demand for a highly educated workforce is so great, that the required number of graduates are not coming through the system to meet the demand.

According to a study undertaken by the Irish Times newspaper in 2016, ``about one-third of Computer Science students across all institutes of technology are dropping out after first year in college" (Carl O'Brien, 2016\nocite{IrishTimes}). Similarly the report discusses high drop-out rates among students progressing from first to second year in universities. It is well accepted that a high contributor to this lower progression rate is that incoming students to CS struggle to master fundamental concepts in their first programming language modules (Quille et al, 2015\nocite{Quille2015}).

What can we do to help solve these problems? There are two things that we believe must be considered. Firstly, we must educate students in the subject area of Computer Science at an earlier age so that they have an inherent interest when it comes to choosing a college/university course. In Ireland steps have been taken at second level to address this. From the beginning of the 2017-2018 school year, Irish secondary schools will begin teaching a short course in coding and other aspects of Computer Science to ``Junior Cycle" students (approximately 12-15 years of age). In 2018, a full Computer Science option to ``Senior Cycle" students (approximately 16-18 years of age) will be offered (Donnelly, 2016\nocite{independent}). Teaching programming at an earlier age is becoming prevalent in many other countries too as the importance of Computer Science becomes more evident. The second thing we need to ensure is that we are teaching students correctly. This means, using the correct methodologies, using the right programming language and starting with the correct basis. All of these are challenges we aim to discuss in this paper, with a particular focus on language choice.

This paper contains the findings of a systematic literature review that was performed between October 2016 and March 2017. In it, two research questions were asked relating to Computer Science retention and what languages/tools we should be using to get the best performance/interest from students of various ages. These questions will help to inform us as to whether visual or textual languages or a hybrid of both is the best choice as a teaching language. It will also determine whether this choice has any bearing on future decisions about (and ability with) programming.

\section{Research Questions}
This study is focused on the relationship between language choice and learning to program. In particular, we want to discover what effects visual programming languages have on the learning process as well as how they compared to the performance of students using traditional text-based languages. To that end, the following research questions were defined: \\

\begin{enumerate}[noitemsep, nolistsep]
	\item Are there any benefits of learning a visual programming language over a traditional text-based language?
	\item Does the choice of First Programming Language make a difference? What languages are the best ones to teach?
\end{enumerate}

\section{Background}
Systematic literature reviews provide an unbiased and comprehensive approach to answering broad research questions. They offer a strict set of guidelines for how to extract information from relevant databases and process it in a detailed manner. This allows for an exhaustive analysis of available papers balanced with time required to process them. For the questions we will raise in this paper, we anticipated a very large amount of material could be found. As such, a literature review was the best approach for us to take.

With Computer Science making its way onto second level school curricula in Ireland, this research is very timely. The topics of language choice and the ``best" First Programming Language and teaching approach is often asked, for example by (\cite{Davies2011}, \cite{Eid2012}, \cite{Ivanovi2015}, \cite{LindaMannilaMichaeldeRaadt2006}, \cite{Quille2015}), but rarely answered. With this review, we aim to address the topics in a detailed manner and compile the opinions and results of many researchers.

Some other literature reviews that have informed this study include the work of \cite{nolan2016role} on anxiety in programming. This well structured review along with Kitchenham's guidelines (2007\nocite{Kitchenham2007}) provided a number of key methodologies for undertaking the review. Some other reviews were also read, but were a lot more specialised. \cite{major2012systematic} looked at teaching introductory programming using robots, for example. Our review covers the full spectrum of introductory programming language opinions and visual/textual language comparisons. It's broad nature will be of great use to researchers and educators alike trying to decide what approaches to use in their classes.

\section{Method}

\subsection{\textbf{Introduction}}
The methodology used to perform this literature review is based on Barbara Kitchenham's approach, as modified by \cite{Kitchenham2007}. This procedure was chosen due to it's high focus on removing human bias from the search process. This ensures, to the highest possible level of certainty, that no false positive answers to the research questions will be found. \\ \\ The method involves performing the following thorough steps, which are followed throughout this paper:

\begin{enumerate}[noitemsep]
	\item Identify the need for a review (See Sections 1 and 3). 
	\item Specify the research questions (See Section 2).
	\item Develop a review protocol (See Sections 4.1 - 4.3).
	\item Identification of research (See Section 4.4).
	\item Study quality assessment (See Section 4.5).
	\item Data extraction and synthesis (See Sections 4.6 - 5.3).
	\item Report on found results (See Section 6).
\end{enumerate}

\subsection{\textbf{Resources Searched}}
Between October 2016 and November 2016, searches were performed on numerous publication databases, namely, the ACM Digital Library, IEEE Xplore, the Education Resources Information Centre (ERIC) and Google Scholar. These particular databases were chosen due to the high level of regard achieved in their respective industries. ACM and IEEE both contain a very wide range of Computer Science papers. ERIC is primarily an educational database, which is also important for this study. Google Scholar was used as a backup database to ensure that all important papers were found.

\subsection{\textbf{Search Terms}}
The methodology used to perform these searches involved taking each primary term and searching for it in each database. If the primary search term alone yielded less than 400 results, all of those paper's were extracted for later filtering. If the search was too broad, it was combined with each respective secondary search term and those results were then chosen for filtering.

Due to the broad nature of this study, an extensive list of search terms was used. This list included 10 primary search terms and 16 secondary search terms. These terms were chosen in order to cover a broad spectrum of age groups and language types. The goal was to be as fully comprehensive as possible. \\

The terms used were: \\

\textbf{Primary Terms:} \textit{``Visual Programming", ``Iconic Programming", ``Visual Versus Textual", ``Visual vs. Textual", ``Graphical Programming", ``Textual Programming", ``First Programming Language", ``Introductory Programming", ``Novice Programmers"} and \textit{``Programming Education"\\}

\textbf{Secondary Terms:} \textit{``Scratch", ``Alice", ``Primary Education", ``National School", ``Elementary Sch\-ool", ``First Level", ``Secondary Education", ``High School", ``Second Level", ``Third Level", ``College", ``University", ``CS1", ``Kids", ``Children", ``Education"} and \textit{``Teaching"}

\subsection{\textbf{Document Selection}}
The initial searches on each database produced a very large number of results. In total (combining the amount of responses for each pair of search terms) ACM returned 2,252 papers, IEEE returned 1,713 papers, ERIC returned 486 papers and Google Scholar returned 655 papers. The first step to minimise these numbers was to perform a ``Title Filtering" on the related papers. This removed any titles where it was immediately obvious they would have nothing to do with the research questions posed. This process cut the number of possible papers down to 661 (all sources merged).

After obtaining full copies of the filtered papers, the next stage involved an ``Abstract Filtering". This was performed in much the same way as ``Title Filtering" but the full abstract of each paper was read. If the content of a paper's abstract did not relate to either of the research questions, it was excluded. This process was undertaken in December 2016. After it's completion, 124 possible papers remained.

At this stage, each paper needed to be read in full. Inclusion and Exclusion Criteria were defined as well as a quality assessment (See Section 4.5) undertaken at the same time during this phase. \vspace{\baselineskip}

The requirements for a paper to be included were that the paper:

\begin{itemize}[noitemsep]
	\item Focused on the topic of at least one research question.
	\item Focused on specific programming languages, either visual, textual, or a combination of both. Specifically, a study/verification needed to be undertaken with students.
	\item Detailed the learning of a First Programming Language.
	\item Was NOT grey literature/blog/a PhD thesis.
	\item Did NOT examine students under the age of 10.
\end{itemize}

Each paper had to meet all applicable of the above requirements. For papers that were borderline author discretion was used based on their content. For example, some papers were kept that described the development of certain Visual Programming Language tools despite not detailing any studies.

\subsection{\textbf{Quality Assessment}}
After reading each paper in full, a final decision was made as to whether it would be included in the results section of this study. For each paper, a rigorous quality assessment protocol was applied. This process was undertaken manually during the reading phase.

Kitchenham (via Keele, 2007\nocite{Kitchenham2007}) lists 18 possible quality assessment questions in her guidelines. For this study, a small subset of four questions were chosen. These were:

\begin{enumerate}[noitemsep]
	\item How credible are the findings?
	\item If credible, are they important?
	\item Is the scope of the study sufficiently wide? (modified)
	\item How well can the route to any conclusions be seen?
\end{enumerate}

For each of these questions and each paper, a score was applied. A score of 1 was given if the paper completely satisfied the question (Y). A score of 0.5 was given if the paper partially satisfied the question (P). A score of 0 was given if the paper failed to satisfy the question (N). Upon first full read through of a paper, these questions were answered. This involved a certain amount of objectivity. For Q1 and Q2, the credibility of the papers had to closely relate to the research questions. For Q3, small studies that contained very little content or had very small experimental groups were excluded. For Q4, it was important that the paper had a logical route to its conclusions and didn’t make any assumptions. This information was all double checked and adjusted where necessary before making the final decision on the included papers.

\begin{figure}
	\centering
	\includegraphics[width=\textwidth,height=520px,keepaspectratio]{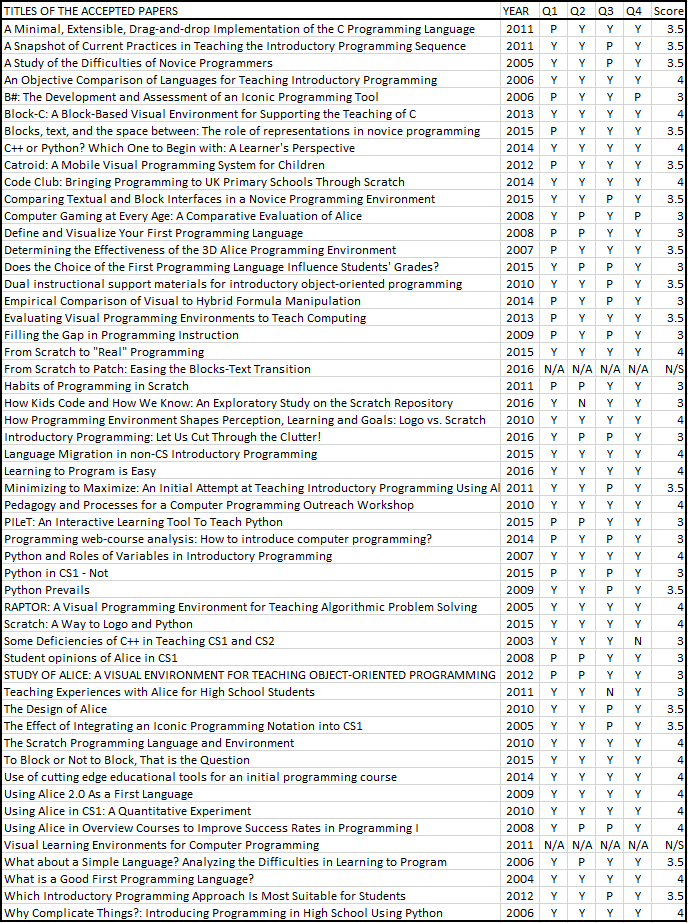}
	\caption{List of accepted papers, their year and their QA Scores (Y=1, P=0.5, N=0, score of 3 out of 4 required)}
	\label{fig:accept}
\end{figure}

For a paper to be included, it must achieve a score of at least three (out of four). This ensures that a paper is of sufficient quality without rashly excluding one that misses a single element. Between this Quality Assessment (QA) assessment and the inclusion criteria, a final list of 53 papers were selected for inclusion. The full list of accepted papers as well as details of their QA scores are presented in Fig.\ref{fig:accept}.

\subsection{\textbf{Data Extraction and Synthesis}}
Throughout the process, all important information was extracted and stored in a number of Microsoft Excel documents. For each search in the initial stages of the study (before coming to the final 53 titles), each individual database search was stored in its own Excel sheet. After title filtering of each document occurred, a master list of titles that passed was created. This document was the primary one used from then on. During the abstract processing stage, papers were highlighted in green if they were to be read in full, highlighted in yellow if they needed further examination and highlighted in red if they were deemed unrelated to the study.

At this point, the 124 abstract filtered papers were split into a new tab on the excel document. Here, the full title, source, publication location, which research question the paper covered and any additional notes were stored. A similar highlighting system was used on this tab as well when papers were deemed to have failed the quality assessment checks or when they did not cover any research questions.

As well as the use of Microsoft Excel, Mendeley reference manager (Mendeley, 2017\nocite{Mendeley}) was used to store every full paper and summaries from the reading of that paper. This tool was chosen as it allowed one to keep track of which papers had been read, and to ``favourite" those ones that passed the quality assessment. A folder structure was used to separate each set of papers into their initial sources (ACM, IEEE, ERIC, Google Scholar). Mendeley also makes it very easy to see where and in what year a paper was published. At the writing stage, Mendeley allowed for easy generation of the list of references for BiBTeX.

\section{Background}

\subsection{\textbf{Types of Studies}}
The vast majority of the included studies involve quantitative experiments involving the results following the teaching of some form of curriculum using a given language. Some authors also used a mixed model approach for data collection (feedback surveys/questionnaires as well as tangible results). Some of the accepted papers were borderline in their Quality Assessment scoring but were still accepted due to the fact that the original developers wrote it, despite not containing any study.

\subsection{\textbf{Timeline of chosen publications}}\label{sect:timeline}
Programming, and in particular programming languages, are a very volatile thing. What may be relevant today might not have been even ten years ago. As such, it was decided to set a hard time-line for acceptable papers. Any paper that passed all other checks and was written any time after 2002 was kept for analysis. This gave a 15 year range for acceptance. This time-line provides a high chance for papers to still be relevant without too many irrelevant studies being kept. Although a lot can change in a 15 year range in terms of Computer Science we felt that in the domain of Computer Science Education there would not be as dramatic a change, as techniques used 15 years ago may still be used today.

\begin{figure}[H]
	\centering
	\includegraphics[width=65mm,height=36mm]{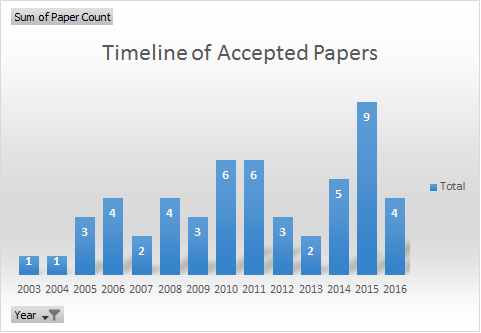} 
	\caption{Timeline of Accepted Papers}
	\label{fig:Timeline}
\end{figure}

The profile of when the accepted papers were published is shown in Fig \ref{fig:Timeline}. A large number of these are from within the last decade. Additionally, 33\% are since 2014. This tends to suggest that our 15 year range is a valid range based on our research questions.

\subsection{\textbf{Data sources}}
\begin{figure}[H]
	\centering
	\includegraphics[width=\textwidth,height=35mm,keepaspectratio]{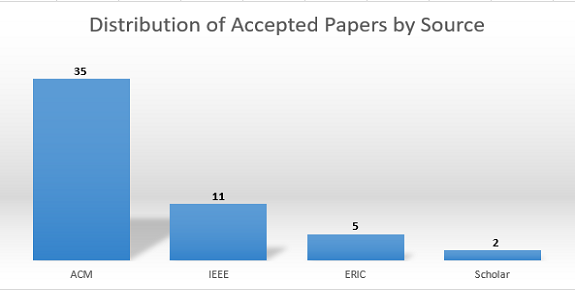}
	\caption{Distribution of Accepted Papers by Source}
	\label{fig:Distribution}
\end{figure}

The largest quantity of accepted papers came from the ACM database. Fig. \ref{fig:Distribution} shows the breakdown of which database the 54 accepted papers were found in. For comparison, during the reading stage (before the final filtering), 73 papers were retrieved from ACM, 36 from IEEE, seven from ERIC and eight from Google Scholar. All accepted papers were disseminated via a conference or a journal.

\subsection{\textbf{Dataset Discussion}}
This study was performed systematically in order to ensure the answers to the research questions were comprehensive, unbiased and valid. As can be seen in Section \ref{sect:timeline}, a 15 year range of acceptable papers was set. This allows for examination of the evolution of teaching methodologies and languages within recent history. With this time-line, we can see what has changed and perhaps more importantly, what has stayed the same. Similarly, with the inclusion of secondary search terms, we are able to look at different levels of education (primary, secondary, tertiary). As will be seen throughout the results, different levels of education tend to converge towards certain language types or teaching styles. This trend appears to be universal. Even though the final papers are from different countries, their results have key similarities. Through this systematic process and with this information in mind, we believe the results returned were of high quality.

\section{Results}
In this section, analysis of the 53 approved papers will be performed. This analysis will involve a second full read-through of each paper (having first read them in the QA stage along with the papers rejected at this stage). While reading the papers, key points were extracted and noted down in the Mendeley Reference Manager notes section. Twenty-nine papers inform the first research question, with 24 informing the second. It is important to note that the level of contribution that some papers will have will be greater than that of others. Some papers that were included may have only raised one strong point, but if it was a point worth making, it was included.

\subsection{\textbf{Research Question 1: Are there any benefits of learning a visual programming language over a traditional text-based language?}}
Before this question is addressed, it is important to first define what is meant by the term ``Visual Programming Language" (VPL). A VPL is any programming language where users are able to manipulate the underlying code in some graphical fashion rather than the traditional text-based approach. Some examples of widely used VPL's today include Scratch (Maloney et al, 2010\nocite{Maloney2010}) and Alice (Cooper, 2010\nocite{Cooper2010}). Before discussing these, let's look at a more traditional approach.

\subsubsection{\textbf{Flowchart Approach}}\label{sect:Flowchart}
A more traditional approach to visual languages came in the form of using flowcharts. Most ``modern" languages don't use this methodology, but for completeness, results from the search that covered this style of design are included here. \cite{Greyling2006} discuss the concept of the B\# language that they developed. B\# uses an iconic flowchart approach to give students two options in developing their code, via drag-and-drop of code pieces, or by the traditional textual approach. As a user is building a flowchart, code is generated in parallel in C++, Pascal or Java. Flow chart icons are connected by lines, making the ordering and structure of the program obvious. While this methodology worked well in the early stages of a CS1 course, the authors note that ``unfortunately initial evaluation sessions showed that many students did not succeed in developing adequate coding skills while working with B\#". Another example of a flowchart based VPL is RAPTOR (Carlisle et al, 2005\nocite{Carlisle2005}). RAPTOR's goal is to improve problem solving skills while reducing the emphasis on syntax. It uses a similar approach to B\#, except without a textual counterpart. Different elements are built up via drag-and-drop, ensuring that program structure is correct. From a study of 959 test subjects, the authors found that students prefer to express their algorithms visually, with 95\% choosing to use a flowchart on the final exam over a textual language. This lends credence to the concept of a VPL, allowing more advanced tools to be built.

\subsubsection{\textbf{Scratch}}
Scratch was developed by the Lifelong Kindergarten Group at the MIT Lab. Scratch's primary goal is to give young people an accessible way to introduce themselves to programming. It uses a ``drag-and-drop" approach, where users drag ``blocks" from a predefined list of commands into a script area. These blocks essentially fit together and make syntax errors impossible. This reduces the mental load of the student and allows them to focus on concepts rather than becoming bogged down with the technicalities of the language.

There have been many studies performed to verify the efficacy of Scratch as a teaching tool for young audiences. \cite{Tangney2010} used Scratch in a project-based after-school workshop for 15-16 year old students. Their goal was to see if they could engage students at an early stage and put them on a path to CS courses. They had 39 students with high maths performance attend the workshops, and the results were favourable. The majority of participants enjoyed the content of the workshop, and the authors noted that ``participants left with a favourable and more realistic impression of both CS courses and the CS profession". 

\cite{Lewis2010} performed a brief comparison study of Scratch and Logo. This comparison focused on programming concepts. Lewis noted that ``the Scratch environment provided a relative improvement in learning outcomes for students learning the construct of conditionals." \cite{Meerbaum-Salant2011} point out however, that bad habits can still happen even in a VPL. They discuss how, if these are not caught, they could actually affect performance in later textual language courses. We as teachers still need to portray good methodologies for students to demonstrate success. 

These papers cover just a small sample of the work that has been done verifying the usefulness of Scratch, and it is well established in its field. Perhaps the clearest indicator of Scratch's success is the sheer number of projects that are connected on their website with 20,695,116  total projects shared when checked in March 2017 (MIT Media Lab, 2017\nocite{ScratchStats}).

\subsubsection{\textbf{Alice}}
Alice, while working in a very similar manner to Scratch has somewhat different targets. The developers (Cooper, 2010\nocite{Cooper2010}) aim was to provide students with a ``serious pre-CS1 programming experience". Alice allows users to build up a functional animated world using drag-and-drop code blocks. It contains a ``Virtual World Editor", which allows users to lay out a set of objects in 3D space. All the underlying code is still dealt with in a drag-and-drop manner after this initial visual setup. Cooper has noted that, ``opposed to algorithm animation, program visualisation systems allow the student to create their own animations". One of Alice's biggest differences from Scratch is that it supports the Object Oriented approach to programming, although in a limited fashion.

Among the studies that investigate Alice as a VPL is the study performed by \cite{Parker}. This study involved a week long workshop with 15 high school students, with a goal of encouraging them towards a degree in Computer Science. The participants did connect with the course, with many stating they enjoyed the video game development aspect and their enthusiasm was encouraging. Larger studies include that of \cite{Sykes2007} which focused on the Objects First approach that Alice allows. A CS1 course based on Alice was directly compared to two iterations of a CS1 course based on C. The author accepts that it is harder to perform computations with Alice, and that the lack of visible syntax could be an issue in later courses, but at the same time, it is noted that Alice makes it very easy to understand the fundamentals, which is exactly what one would want from a CS1 language. The Alice students outperformed the control groups significantly in the exams. 

\cite{Johnsgard2008} present another success story using Alice. They were experiencing low grades in their C++ based CS1 course. They implemented a CS0 using Alice. The average grade in the following year of C++ rose to 70.3\% from 46.4\% in the previous year, a statistically significant increase. Student's also expressed their enjoyment of the Alice course. \cite{Anniroot2012} found that Alice helped their students better their problem solving abilities and gave them a stronger understanding of programming concepts. They found that through ``quantitative analysis of the closed-ended questions, 81\% of experimental learners were found to agree that the visual effects in Alice provide meaningful contexts for understanding classes, objects, methods, and events". 

Of course, not everyone can have a positive experience. \cite{Garlick2010} felt that while Alice was a nice tool, students didn't necessarily focus on programming concepts enough and were more just enjoying building a world. The alternative they offered was a ``pseudo code" CS0. In other words, a course that focused purely on algorithmic design with no programming language used at all. The participants in the course had similar results in Alice and pseudo code assignments, with a worse result on the Alice exam, and the Alice group declared less confidence as well via collected survey responses. Given the large number of positive feelings towards Alice in multiple studies (a small sample of which are discussed here), Alice's failings in this course could possibly be attributed to the teaching techniques employed or the fact that pseudo code might not be comparable to a full language (i.e. programming is more complex).

\subsubsection{\textbf{After school Clubs}}
After school clubs, such as CoderDojo, often teach Scratch or a similar VPL as one of their modules. The authors have had the pleasure to watch some students progress from a local CoderDojo to the CS1 course on offer at their university. These students often have a strong advantage and perform very highly at college level. 

In 2014, \cite{Smith2014} analysed the effect of 1000 Code Club locations in UK schools. They chose to primarily teach Scratch due to its ``known ease of use by primary school children". Step by step instructions were given at first, but as the children progressed, they were expected to create their own scripts and make their own choices. Surveys were collected at the end of the year, with some positive results. The children had demonstrated a good knowledge of some programming concepts. Smith analysed 22 final projects at random and discovered that ``(some) children coped remarkably easily with difficult programming concepts". To push this idea further, \cite{Seals2008} had eight to nine year old's working on assignments in Alice that were of an equivalent difficulty level to that which 18-19 year old college students would undertake. This is quite remarkable. Something must be helping these young people understand things so clearly.

\subsubsection{\textbf{Blocks}}\label{sect:blocks}
The most prevalent thing noticeable in the analysis of both Alice and Scratch is that the block-based approach to teaching seems to resonate strongly with younger cohorts. This may not be a surprise due to the fact that a large number of people are believed to be visual learners, and young students generally have more creative minds. What else is it that makes this blocks approach so strong?

\cite{Sandoval-Reyes2011} asked themselves this same question. They performed an analysis of three major block programming environments: Scratch, Alice and App Inventor, while also looking at Greenfoot. They put forward the idea that this kind of environment provides such strong pedagogy due to ``connecting users with their interests", direct mapping of ideas to instructions on screen and the hiding of unnecessary complexities from the novice user. The blocks approach can work in all kinds of environments, as demonstrated by Catroid (Slany, 2012\nocite{Slany2012}). Catroid allows users to develop programs directly on their phone, they can even develop controllers for other devices such as the Lego Mindstorms NXT robot. For many young people, this is a really exciting prospect. \cite{Price2015} undertook an interesting study that ``seeks to isolate the effect of a block interface on the experience of novices". Half of a group of middle school students were assigned to a ``block" group, while the other half were assigned to a ``text" group. During a half-day session, the students were given programming exercises to do in their respective environments. At the end of the session, data via surveys and logged interactions with the tools were analysed. The block group performed better than the text group, and they also had a slightly higher self-efficacy at the end of the session as well. This provides strong data pointing to blocks having some positive effect on the performance of students.

\subsubsection{\textbf{Transition from Visual to Textual}}\label{sect:transition}
A number of researchers tend to agree that, while Visual Programming is a very strong concept for introductory courses, it has a tendency to fall short when the time comes to deal with complex topics. Some researchers agree that VPL's are more of a ``gateway" to learning textual languages.

\cite{Dorling2015}, for example, examined a scenario in which graphical languages were taught ``in conjunction with, not in place of, text-based programming languages". This study involved beginning a 10 week curriculum using Scratch and algorithmic concepts, and working towards introducing Python. By showing students Python code side-by-side with Scratch code, their understanding of the textual language was made much stronger. It was noted that ``this transition process has been a factor in an increased uptake of Computer Science". \cite{Giordano2014} looked at a similar approach that involved using multiple languages in the same course. This study was done with a group of 28 10\textsuperscript{th} grade students in Italy between the ages of 14 and 16. The course was taught over 28 weeks. The early stages used Scratch and similar tools. According to the authors, ``This is done in order to relieve students from the burden of learning all the syntax-related details and instead to let them focus on the concepts and problem solving skills." In later weeks, the C language was introduced to give students a proper experience with a textual language. The results were positive across the course. In particular, when the C language section began, students made less errors than would normally be expected upon first exposure. This shows that using the VPL first has allowed students to familiarise themselves with the concept of programming. \cite{Weintrop2015} asked ``To Block or not to Block". They wanted to determine if high school students found the blocks approach easier than the textual approach and why. To examine this, they taught a course using five weeks of Snap! and five weeks of Java. Fifty-eight percent of students found Snap! easier to use. Some participants reported (via a survey) that the blocks approach was easier to read and the shapes were also determined to be helpful. There are also some drawbacks, for example, blocks languages are less powerful. At some point, you will hit a barrier you can't pass with the tools you have. The author suggests an interesting point: ``Why not add a similar browsability to introductory text-based environments?". Multiple authors have examined this concept in detail, we will call this type of language a "Hybrid Language".

\subsubsection{\textbf{Hybrid Languages}}
These languages involve either an interface that shows both visual and textual elements at the same time, or the merging of a textual language into a blocks style interface. \cite{Sciences2015} describes this idea in detail. A pilot study involving the use of blocks-based, text-based and hybrid programming environments was performed in order to compare the effects of all three. The analysis (while ongoing) showed promise for the concept of hybrid languages. There are certainly benefits to both visual and textual approaches, hence why combining them might have the best possible effect on young learners.

The earliest found study on the concept of hybrid languages was undertaken by \cite{Cilliers2005}. The authors wanted to examine what effect the integration of an iconic notation into a textual development environment would have. They recognised that ``visual programming notations offer benefits over textual programming notations" while also recognising that VPL's were not a standalone solution. In order to verify their thoughts, they implemented a course that compared a control group using exclusively PASCAL as their language of choice to a study group using B\# as their language of choice (first mentioned in Section \ref{sect:Flowchart}). B\# was designed with the intention of only being valid for the initial stages of CS1, after which the students would progress to a purely textual approach (once they became familiar with the concepts). Participants using B\# performed statistically significantly better upon final assessment, particularly amongst students deemed to be high risk. In other words, it helped those who would have struggled quite a lot with the traditional approach, without having a negative effect on those who didn't necessarily need the extra help. 

\cite{Koitz2014} also asked a similar question when comparing Scratch and ``Pocket Code". Pocket Code is a mobile development environment that uses a mix of textual programming and Scratch elements. After performing four tasks using both languages (17 participants), the results showed that Pocket Code's hybrid approach was more beneficial than the purely visual approach of Scratch.

In recent years, there has been a large amount of research into hybrid languages that use existing textual languages merged with a block style model. In 2011, \cite{Federici2011} combined the C programming language into a Scratch-like blocks system. This was made possible by a Scratch mod known as BYOB. This tool allows for the creation of custom blocks and functions within the Scratch environment. The author implemented blocks such as printf, scanf, integers etc. The goal of this research was to ``lower the student effort required in advancing from introductory tools, such as Scratch, to regular programming languages, such as C". The tool they designed was named blockC. \cite{Charalampos2015} developed a very similar implementation called Block-C. The authors wanted a visual methodology without skewing away from teaching a ``general purpose programming language". This was tested with a two hour tutorial and 32 first year university students. The Block-C group performed much stronger than a textual C group.

This concept expands past just the C language of course, with tools existing to allow blocks to be modified for any language. \cite{Matsuzawa2015} provide an example of a Java based blocks language. With their study, the tool allowed for direct translation between blocks and text-based Java. The author posited that students should begin with this blocks approach, and gradually move towards a fully textual environment. Those who continued to use Blocks for the entirety of the study turned out to be the weaker students, showing that the visual approach might have a threshold. Finally, \cite{Robinson2016} looked at a tool called ``Patch" which combined Scratch with elements of Python. Again, the goal was to minimise the gap between visual and textual languages in young learners. No verification was done on this particular tool, but it follows much the same patterns as the other tools discussed.

\subsubsection{\textbf{Academic Benefit}}\label{sect:benefits}
Based on all of the discussed approaches to visual programming, can we see any academic benefit to teaching such a language? It is well established that younger students can take a tool like Scratch and really thrive using it, but what about second level students? \cite{Cheung2009} believes that there are key age groups for each type of language to be most successful. High school students respond better to textual programming, students younger than 14 find VPL's the most beneficial and those from around 15-17 would most benefit from a hybrid environment. They ran summer workshops that back up this fact on hybrid languages. \cite{Andujar2013} also wanted to see if there was any benefit of teaching high school students visual programming. They came to a similar conclusion as \cite{Cheung2009}, in that teaching Alice to them did not provide any significant benefit over other languages. However, Alice did increase the retention rate of students. This could come down to the enjoyment of using a VPL, and this effect could hold true for all courses using VPL's.

\subsubsection{\textbf{Conclusion}}
From the literature, it is clear that Visual Programming Languages present many benefits over traditional text based programming languages. As presented in Section \ref{sect:benefits}, all types of language have their benefits. We see in Section \ref{sect:blocks}. that middle school students both performed better and had higher self-efficacy when using the blocks based approach to programming (Price and Barnes, 2015). In Section \ref{sect:transition}, multiple examples are provided showcasing the effects learning a VPL can have on learners as they progress towards a TPL. Having this knowledge and skill is a key factor. After school clubs such as CoderDojo encourage young kids to get involved with programming at an early age. From our own perspective, we have seen multiple students progress to our CS1 course and perform at the top of the class. For the purpose of this research question, we can certainly conclude that teaching a Visual Programming Language or hybrid programming language to the right age group can have a very positive effect on their interest and retention in Computer Science. This might have a positive effect on the retention rates in college level courses.

\subsection{\textbf{Research Question 2: Does the choice of First Programming Language make a difference? What languages are the best ones to teach?}}
In this section, we will examine whether or not the choice of First Programming Language (FPL) has a significant effect on outcome in an introductory programming course. Specific languages will be examined, and conclusions will be drawn.

\subsubsection{\textbf{A ``Good" First Programming Language}}\label{sect:good}
The first question that must be asked is, what constitutes a good FPL? \cite{Gupta2004} examined this question in detail and believes that the choice of FPL is a big decision, one that will have a ``profound impact" on future learning. He concluded that the ``ideal" language will depend on the age of the target audience among other things. He posits that it is important to ``focus on problem based learning, allowing students to focus on techniques rather than on the language syntax itself". Some of the important elements of a FPL that he discusses are:

\begin{itemize}[noitemsep]
	\item The language should have a clear and intuitive syntax.
	\item The language should cover all common syntactic and semantic constructs.
	\item The language should be consistent in it's handling of things like errors and provide meaningful error messages.
	\item The language shouldn't have excess brevity (functional languages) or excess verbosity (Java boilerplate).
	\item The language should be customisable and allow for changing needs over time.
\end{itemize}

\cite{Ateeq2014} examined this research question specifically in the context of C++ or Python. They agree with many of Gupta's definitions of important features in a FPL, particularly regarding notation overhead, verbosity, target audience and use of simple syntax. They found that Python met many of these requirements (which will be discussed in more detail in Section \ref{sect:textualFPL}). To test this, a study was run with CS1 students comparing their thoughts of both languages (via surveys). In most regards, Python was held in a higher regard which further proves that the checklist above is an important factor. \cite{LindaMannilaMichaeldeRaadt2006} also examined objectively what languages might be the best to use as a FPL. A list of 17 criterion was developed by educational language writers. Eleven well known languages were examined against these criterion. Those that came out on top were Python (meeting 15 of 17 criterion), Eiffel (15/17) and Java (14/17).

\cite{Ranade2016} furthers these ideas by talking about his college's use of a C++ language that has been graphically augmented with a logo turtle style view. He believes that the focus of a FPL course should be taken away from syntax/semantics and directed towards the more fun aspects of computing as well as algorithmic thinking. One key example from their work was the use of this tool to demonstrate how recursion works using a visual tree that keeps splitting its branches in two as the structure grows deeper. This tree was drawn in real time. The author believes the visual nature of this allows for easier comprehension of complex concepts.

\subsubsection{\textbf{Difficulties with CS1}}\label{sect:difficulties}
The attrition rates in CS1 courses are often quite high; there must be some attributing factors to this. \cite{Lahtinen2005} used a survey to help discover what some of the key difficulties students experienced were. This survey was distributed to 559 students and 34 teachers from a group of multinational universities and colleges. Most were students of C++, but some had used others as well. Most of the key results agree with the ideas of a good FPL in Section \ref{sect:good}. Recursion, pointers, abstract types and error handling were determined to be the hardest concepts. Getting familiar with structures, syntax, algorithm design and how to divide into functions/classes are the elements that need to be done to be successful. In general however, ``the teaching language did not seem to affect the learning situations". \cite{Mannila2006} also analysed Java and Python programs with the intent of determining the difficulties the writers experienced. Sixty programs written by 16-19 year old novices were used. Common errors that were found involved poor error checking, bad use of variable types in Java and mismatching brackets in Java. The authors also agreed that Python had the potential to be a strong CS1 as it had less errors than the verbose Java code.

\cite{Luxton-Reilly2016} on the other hand posits that learning to program is actually an easy endeavour, and that we, as educators, expect too much from students in a CS1 module. Could we in fact be scaring people away from CS by overestimating how much can be learned in a short period? He raises the point that ``There is nothing intrinsic to the subject that makes it difficult to learn, but rather our subjective assessment of how much a student ``should" be able to achieve by the end of the course that determines the difficulty". This is something to keep in mind as we discuss FPL's next.

\subsubsection{\textbf{The commonly chosen languages}}\label{sect:commonly}
\cite{Davies2011} conducted a survey of 371 institutions in the US in 2011. This will give a reasonable snapshot of FPL choice in general in this region. They broke the survey down into CS0, CS1 and CS2. The most commonly used CS0 language was Alice, followed by Python and Java. For CS1, the primary focus for this paper, Java was the most used with 48.2\% of institutions adopting it, 28.8\% offered C++ and 12.9\% offering Python. Alice only maintained 4.3\% usage as a CS1 language. For CS2, the usage of Java strengthened further to 55.8\%, with C++ also increasing to 36.1\%. All other languages of note fell to usage rates of below 4\%. This tells us that advanced topics are much better suited to object oriented environments. In the following sections, the efficacy of a subset of these languages will be discussed in detail.

\subsubsection{\textbf{Textual FPL's}}\label{sect:textualFPL}
By far, the most commonly found language in the literature was Python. This may come down to the fact that Python is reasonably new when compared to C++ and Java. These languages have already cemented themselves in the pedagogy of CS1. Python still needs to convince educators of its efficacy, however it could be a strong choice for a FPL.

\cite{Grandell2006} discuss their attempts at a Python based FPL course with high school students. They recognised that Python met a lot of the requirements that make it ``easy" to learn. They implemented a curriculum and tested it on 42 boys. Eighty-five percent of students passed the course, with an average grade of 77.1\%. This was compared to a similar High School course in Java they previously taught, with the Python average being much higher. Survey results showed that students strongly agreed that Python was easy to learn. \cite{Nikula2007} also considered Python to be an easy language to learn. They tested this at three institutions that previously used different languages (C, Java, Delphi). In all cases, Python was found to be a better choice. This was determined by a higher average grade on a course of comparable difficulty. \cite{Leping2009} used Python as their FPL with a subset of their class in 2008. The rest of the class was still taught using Java. They felt that Python was ``elegant, simple and practical" with clean and easy to read syntax. The results showed similar outcomes for both Java and Python students. One interesting outcome however was that a lower percentage of people outright failed the course in Python, but more students were perhaps scraping by.

\cite{Hunt2015} disagrees with \cite{Grandell2006}, \cite{Nikula2007} and \cite{Leping2009}. In 2014, his department switched from teaching Java to Python. In 2015, they decided to switch back due to problems they experienced that hadn't been noted in literature studied. In particular, the lack of arrays, the difficulty of transitioning to Java in CS2, and the inability to focus on an ``objects first" approach were cited as the reasons for this.

According to the TIOBE index (Software, 2017\nocite{TIOBE}), Java is the most commonly used programming language in the world, so it makes sense that it is a frequently used FPL also. \cite{Ivanovi2015} took up teaching of Java after a number of years of using Modula-2. They decided to do a comparison study. While they liked Java as a language, there were no statistically significant differences in grades. The author posits ``this result suggests that the choice of the introductory programming language does not matter if we use students' performance as the criterion of suitability". Again, it is worth noting that many other papers in the literature mentioned Java as their FPL without discussing why. This could be complacency due to it being such a widely used language that not many researchers are discussing its efficacy.

Not much information was found relating to C++ as a FPL. This is likely due to the rise in Java in the last two  decades. One paper that discusses C++ was \cite{Agarwal2003} which covers some of the issues related to C++ in CS1. It is noted that C++ contains many verbose and over-complicated elements such as \textit{include statements, unnecessary typecasting and string comparison}. A lot of these issues are also present in other common FPL's. The author is not trying to discourage the use of C++, but merely pointing out some likely pitfalls that could be experienced.

\subsubsection{\textbf{Visual FPL's}}
Scratch is commonly used to teach  programming to young students, but is it effective? \cite{Aivaloglou} performed an analysis on a database of 250,166 scraped Scratch projects to see how children make use of the tool. While most projects were small, conditionals and variables were frequently applied. There were also some example of large projects using multiple sprites and many blocks of code. This shows that Scratch has the potential to be used as a FPL, or in general as a first-exposure programming environment. They also found a high count of clone's within the data, suggesting that it already is being highly used for teaching purposes. \cite{Armoni2015} noted that learning Scratch at an early age did affect retention. These students chose to continue on to a Java/C\# course later into their school lives. They also appeared to pick up information faster and grasp the tougher concepts before their peers.

As presented in Section \ref{sect:commonly}, Alice is the most frequently used language in college level CS0 courses. \cite{Mullins2009} discussed one such course. The authors see the importance of students being able to see and manipulate objects directly in the editor, a benefit they would not experience in a textual language. Upon examining collected data from the course, it was noted that results varied, but for the most part using Alice increased pass rates, sometimes with a lower average grade however. In general, Alice proved most helpful for those students who would traditionally struggle with the material, without having a negative effect on other students who don't need the extra help. Retention and interest also increased with those who undertook this course.

\subsubsection{\textbf{Comparison of Textual and Visual FPL's}}
A multitude of studies have been performed that compare the usage of VPL's and textual languages as FPL's. \cite{daprogramming} wanted to determine if, and how, VPL's can help learners understand and transition to a textual language. This was tested with the help of a pair of Moodle based web courses. The visual course was taught using Visual iVProg and the textual course used C. The Moodle ``Virtual Programming Lab" environment allowed for automatic evaluation of student submissions. These courses were voluntary (open to the public), and only lasted four weeks. There was a total of 144 participants, split between both teaching styles. The content in both courses was essentially the same. From an analysis of workload, the authors conclude that ``visual programming seems to be a nice option to introduce programming concepts".

\cite{Cliburn2008} discusses a CS1 course that taught Alice and Java together in the same term. A total of 84 participants took this course, of which 59.5\% found Alice helped them in their understanding of Java. Despite this, the author argues that this outcome was not good enough. If Alice truly made a lasting difference, apart from just the effect of knowing elements of a programming language before beginning Java, then the figure should be much higher. One interesting response from a provided survey was ``the programming concepts it (Alice) taught were mostly so simplistic that it really would have been better to spend only a little time on them and the more complex concepts did not make sense until I learned them in Java". It was due to responses like this that the author decided to revert to a full Java course. He still believes that Alice can be useful, but perhaps not in a two language, one semester style course.

\cite{Daly2011} also compared the effects of teaching Alice side-by-side with Java. The focus was on confidence levels and if they have an effect. There were a total of 29 participants who took part in this online study. Eighteen took the pure Java course, with 11 taking a course that entailed six weeks of Alice followed by six weeks of Java. The author found that ``the students in the Alice/Java course had a higher level of confidence overall when compared to the pure Java course". More importantly, confidence did seem to imply success in the course and also led to higher retention and enjoyment.

\cite{Eid2012} wanted to investigate if learning introductory concepts in a textual language was better than doing so in a VPL. Two groups of students were examined. One group started with a textual language and proceeded to a high level visual programming course. The other group started with a low level visual programming course and proceeded to the high level one. This allows for the focus on concepts first and lets students understand what's happening at a basic level. The authors found that there was statistically significantly better test results for those whose FPL was text-based.

\subsubsection{\textbf{Textual Augmentation}}
A number of authors also look\-ed at the concept of textual augmentation, which is akin to the hybrid languages discussed earlier. \cite{Laakso2008} wanted to look at the concept of an executable pseudo language. The authors believe that this allows you to take the focus away from verbose syntax while still allowing the run time nature of programming to shine through. Their solution involved a tool called ViLLE, which runs on a subset of Python, and allows for visualisation. The authors tested this tool on a class of 72, with 32 students using ViLLE. The results showed enhanced learning in those who used ViLLE.

\cite{Montero2010} looked at Greenfoot which allows for visualisation of object oriented Java concepts using animation. They chose Greenfoot as it allowed for both visual and textual editing of the program. In their study, 15 students used the Greenfoot environment while 18 used only textual materials. There was a statistically significant difference at the end of the course in the knowledge of Greenfoot students versus the knowledge of the control group about Object Oriented principles. Greenfoot was also liked by the students, which is always a positive thing.

\cite{Alshaigy2015} developed a tool called PILeT, which is an interactive learning tool for Python. The goal of this tool was to be adaptable to the learning style of the student. If they were a visual learner, they could use a visual tool, similarly a textual model and a puzzle based model were included. As you use the tool, it builds up a knowledge database about how you learn in order to present the user with the best material first time as they progress. The authors goal is to avoid a single pedagogical learning style, which would not necessarily meet everyone's needs. Based on the literature in this paper, and the amount of different approaches that different institutions take, this might be a very strong concept. The analysis of this tool is still ongoing, but early results are promising.

\subsubsection{\textbf{Conclusion}}
It is clear that the choice of a first programming language does not often matter, within reason. Many researchers mention the fact that using correct methodologies independent of what particular tool you are using is the most important thing. Educators have had success with a broad range of different FPL's, and equally others have had failings with many languages. As discussed in Section \ref{sect:good}, there are certain criteria that a ``Good" FPL might have. If these guidelines are followed (by not picking an overly ``difficult" language), along with if the teacher is familiar with a given language, this might lead to the best quality of course. For us, if one particular language of each type had to be chosen however, Python seems to be the most highly supported textual language from the literature, possibly due to its relative newness as a programming language. Java would also be a strong choice as it is currently the most used language in the world (Software, 2017\nocite{TIOBE}) and has proven itself to be a strong FPL (Mannila and de Raadt, 2006\nocite{LindaMannilaMichaeldeRaadt2006}). Scratch is also held in high regard as a VPL. Based on the knowledge that hybrid languages provide the best of both worlds, the ``ideal" language choice might by a combination of Python (or Java) and Scratch. If a course can be made stimulating and interesting for the students then the choice of programming language is not as important as many people think. 

\section{Discussion}
Throughout this review, we have discussed the benefits of learning a Visual Programming Language and whether or not the First Programming Language choice has profound effect on student performance and interest. It is clear that the most important thing educators can do is make their course interesting, and ensure it covers all the important elements needed to truly ``know" programming. If a student of any age enjoys what they are doing, there is a better chance that they are going to understand it and continue studying it.

It has been demonstrated through the answers to the research questions that the actual choice of what tools to use does not matter, within reason. The use of a Visual Programming Language will in most cases, be very helpful to a student. It may not be something to pursue for a longitudinal time frame, but as an introduction to CS, it is clearly beneficial and will generally lead to higher retention of knowledge and interest.

We present this paper in the hope that educators at all levels and in all institution types will examine the options available to them when they are teaching programming. This review may go some way to informing their decisions around the first programming language to use and the benefits of both text-based and visual-based programming languages.

\bibliographystyle{spbasic}
\bibliography{VisTex}

\begin{thebibliography}{62}
\providecommand{\natexlab}[1]{#1}
\providecommand{\url}[1]{{#1}}
\providecommand{\urlprefix}{URL }
\expandafter\ifx\csname urlstyle\endcsname\relax
  \providecommand{\doi}[1]{DOI~\discretionary{}{}{}#1}\else
  \providecommand{\doi}{DOI~\discretionary{}{}{}\begingroup
  \urlstyle{rm}\Url}\fi
\providecommand{\eprint}[2][]{\url{#2}}

\bibitem[{Aivaloglou and Hermans(2016)}]{Aivaloglou}
Aivaloglou E, Hermans F (2016) How kids code and how we know: An exploratory
  study on the scratch repository. In: Proceedings of the 2016 ACM Conference
  on International Computing Education Research, ACM, pp 53--61

\bibitem[{Alshaigy et~al(2015)Alshaigy, Kamal, Mitchell, Martin, and
  Aldea}]{Alshaigy2015}
Alshaigy B, Kamal S, Mitchell F, Martin C, Aldea A (2015) Pilet: an interactive
  learning tool to teach python. In: Proceedings of the Workshop in Primary and
  Secondary Computing Education, ACM, pp 76--79

\bibitem[{Andujar et~al(2013)Andujar, Jimenez, Shah, and
  Morreale}]{Andujar2013}
Andujar M, Jimenez L, Shah J, Morreale P (2013) Evaluating visual programming
  environments to teach computing to minority high school students. Journal of
  Computing Sciences in Colleges 29(2):140--148

\bibitem[{Anniroot and de~Villiers(2012)}]{Anniroot2012}
Anniroot J, de~Villiers MR (2012) {A study of Alice: A visual environment for
  teaching object-oriented programming}. Proceedings of the IADIS International
  Conference on Information Systems 2012

\bibitem[{Armoni et~al(2015)Armoni, Meerbaum-Salant, and Ben-Ari}]{Armoni2015}
Armoni M, Meerbaum-Salant O, Ben-Ari M (2015) From scratch to “real”
  programming. ACM Transactions on Computing Education (TOCE) 14(4):25

\bibitem[{Ateeq et~al(2014)Ateeq, Habib, Umer, and Rehman}]{Ateeq2014}
Ateeq M, Habib H, Umer A, Rehman MU (2014) C++ or python? which one to begin
  with: A learner's perspective. In: Teaching and Learning in Computing and
  Engineering (LaTiCE), 2014 International Conference on, IEEE, pp 64--69

\bibitem[{Bergin et~al(2003)Bergin, Agarwal, and Agarwal}]{Agarwal2003}
Bergin J, Agarwal A, Agarwal K (2003) Some deficiencies of c++ in teaching cs1
  and cs2. ACM SIGPlan Notices 38(6):9--13

\bibitem[{Carl~O'Brien(2016)}]{IrishTimes}
Carl~O'Brien NIM Joe~Humphreys (2016) {Concern over drop-out rates in computer
  science courses}.
  \urlprefix\url{http://www.irishtimes.com/news/education/concern-over-drop-out-rates-in-computer-science-courses-1.2491751},
  [Online; accessed 26-January-2017]

\bibitem[{Carlisle et~al(2005)Carlisle, Wilson, Humphries, and
  Hadfield}]{Carlisle2005}
Carlisle MC, Wilson TA, Humphries JW, Hadfield SM (2005) Raptor: a visual
  programming environment for teaching algorithmic problem solving. Acm Sigcse
  Bulletin 37(1):176--180

\bibitem[{Cheung et~al(2009)Cheung, Ngai, Chan, and Lau}]{Cheung2009}
Cheung JC, Ngai G, Chan SC, Lau WW (2009) Filling the gap in programming
  instruction: a text-enhanced graphical programming environment for junior
  high students. In: ACM SIGCSE Bulletin, ACM, vol~41, pp 276--280

\bibitem[{Cilliers et~al(2005)Cilliers, Calitz, and Greyling}]{Cilliers2005}
Cilliers C, Calitz A, Greyling J (2005) The effect of integrating an iconic
  programming notation into cs1. In: ACM SIGCSE Bulletin, ACM, vol~37, pp
  108--112

\bibitem[{Cliburn(2008)}]{Cliburn2008}
Cliburn DC (2008) Student opinions of alice in cs1. In: Frontiers in Education
  Conference, 2008. FIE 2008. 38th Annual, IEEE, pp T3B--1

\bibitem[{Cooper(2010)}]{Cooper2010}
Cooper S (2010) The design of alice. ACM Transactions on Computing Education
  (TOCE) 10(4):15

\bibitem[{Daly(2011)}]{Daly2011}
Daly T (2011) Minimizing to maximize: an initial attempt at teaching
  introductory programming using alice. Journal of Computing Sciences in
  Colleges 26(5):23--30

\bibitem[{Davies et~al(2011)Davies, Polack-Wahl, and Anewalt}]{Davies2011}
Davies S, Polack-Wahl JA, Anewalt K (2011) A snapshot of current practices in
  teaching the introductory programming sequence. In: Proceedings of the 42nd
  ACM technical symposium on Computer science education, ACM, pp 625--630

\bibitem[{Donnelly(2016)}]{independent}
Donnelly K (2016) {Computer Science finally on the way for Leaving Cert
  students}.
  \urlprefix\url{http://www.independent.ie/irish-news/education/computer-science-finally-on-the-way-for-leaving-cert-students-34576921.html},
  [Online; accessed 14-March-2017]

\bibitem[{Dorling and White(2015)}]{Dorling2015}
Dorling M, White D (2015) Scratch: A way to logo and python. In: Proceedings of
  the 46th ACM Technical Symposium on Computer Science Education, ACM, pp
  191--196

\bibitem[{Eid and Millham(2012)}]{Eid2012}
Eid C, Millham R (2012) Which introductory programming approach is most
  suitable for students: Procedural or visual programming? American Journal of
  Business Education (Online) 5(2):173

\bibitem[{Federici(2011)}]{Federici2011}
Federici S (2011) A minimal, extensible, drag-and-drop implementation of the c
  programming language. In: Proceedings of the 2011 conference on Information
  technology education, ACM, pp 191--196

\bibitem[{Garlick and Cankaya(2010)}]{Garlick2010}
Garlick R, Cankaya EC (2010) Using alice in cs1: a quantitative experiment. In:
  Proceedings of the fifteenth annual conference on Innovation and technology
  in computer science education, ACM, pp 165--168

\bibitem[{Giordano and Maiorana(2014)}]{Giordano2014}
Giordano D, Maiorana F (2014) Use of cutting edge educational tools for an
  initial programming course. In: Global Engineering Education Conference
  (EDUCON), 2014 IEEE, IEEE, pp 556--563

\bibitem[{Grandell et~al(2006)Grandell, Peltom{\"a}ki, Back, and
  Salakoski}]{Grandell2006}
Grandell L, Peltom{\"a}ki M, Back RJ, Salakoski T (2006) Why complicate
  things?: introducing programming in high school using python. In: Proceedings
  of the 8th Australasian Conference on Computing Education-Volume 52,
  Australian Computer Society, Inc., pp 71--80

\bibitem[{Greyling et~al(2006)Greyling, Cilliers, and Calitz}]{Greyling2006}
Greyling J, Cilliers C, Calitz A (2006) B\#: The development and assessment of
  an iconic programming tool for novice programmers. In: Information Technology
  Based Higher Education and Training, 2006. ITHET'06. 7th International
  Conference on, IEEE, pp 367--375

\bibitem[{Gupta(2004)}]{Gupta2004}
Gupta D (2004) What is a good first programming language? Crossroads 10(4):7--7

\bibitem[{Hunt(2015)}]{Hunt2015}
Hunt JM (2015) Python in cs1-not. Journal of Computing Sciences in Colleges
  31(2):172--179

\bibitem[{Ivanovi{\'c} et~al(2015)Ivanovi{\'c}, Budimac, Radovanovi{\'c}, and
  Savi{\'c}}]{Ivanovi2015}
Ivanovi{\'c} M, Budimac Z, Radovanovi{\'c} M, Savi{\'c} M (2015) Does the
  choice of the first programming language influence students' grades? In:
  Proceedings of the 16th International Conference on Computer Systems and
  Technologies, ACM, pp 305--312

\bibitem[{Johnsgard and McDonald(2008)}]{Johnsgard2008}
Johnsgard K, McDonald J (2008) Using alice in overview courses to improve
  success rates in programming i. In: Software Engineering Education and
  Training, 2008. CSEET'08. IEEE 21st Conference on, IEEE, pp 129--136

\bibitem[{Keele(2007)}]{Kitchenham2007}
Keele S (2007) Guidelines for performing systematic literature reviews in
  software engineering. In: Technical report, Ver. 2.3 EBSE Technical Report.
  EBSE, sn

\bibitem[{Koitz and Slany(2014)}]{Koitz2014}
Koitz R, Slany W (2014) Empirical comparison of visual to hybrid formula
  manipulation in educational programming languages for teenagers. In:
  Proceedings of the 5th Workshop on Evaluation and Usability of Programming
  Languages and Tools, ACM, pp 21--30

\bibitem[{Kyfonidis et~al(2015)Kyfonidis, Moumoutzis, and
  Christodoulakis}]{Charalampos2015}
Kyfonidis C, Moumoutzis N, Christodoulakis S (2015) Block-c: A block-based
  visual environment for supporting the teaching of c programming language to
  novices

\bibitem[{Laakso et~al(2008)Laakso, Kaila, Rajala, and Salakoski}]{Laakso2008}
Laakso MJ, Kaila E, Rajala T, Salakoski T (2008) Define and visualize your
  first programming language. In: Advanced Learning Technologies, 2008.
  ICALT'08. Eighth IEEE International Conference on, IEEE, p 324

\bibitem[{Lahtinen et~al(2005)Lahtinen, Ala-Mutka, and
  J{\"a}rvinen}]{Lahtinen2005}
Lahtinen E, Ala-Mutka K, J{\"a}rvinen HM (2005) A study of the difficulties of
  novice programmers. In: Acm Sigcse Bulletin, ACM, vol~37, pp 14--18

\bibitem[{Leping et~al(2009)Leping, Lepp, Niitsoo, T{\~o}nisson, Vene, and
  Villems}]{Leping2009}
Leping V, Lepp M, Niitsoo M, T{\~o}nisson E, Vene V, Villems A (2009) Python
  prevails. In: Proceedings of the International Conference on Computer Systems
  and Technologies and Workshop for PhD Students in Computing, ACM, p~87

\bibitem[{Lewis(2010)}]{Lewis2010}
Lewis CM (2010) How programming environment shapes perception, learning and
  goals: logo vs. scratch. In: Proceedings of the 41st ACM technical symposium
  on Computer science education, ACM, pp 346--350

\bibitem[{Luxton-Reilly(2016)}]{Luxton-Reilly2016}
Luxton-Reilly A (2016) Learning to program is easy. In: Proceedings of the 2016
  ACM Conference on Innovation and Technology in Computer Science Education,
  ACM, pp 284--289

\bibitem[{Major et~al(2012)Major, Kyriacou, and Brereton}]{major2012systematic}
Major L, Kyriacou T, Brereton OP (2012) Systematic literature review: teaching
  novices programming using robots. IET software 6(6):502--513

\bibitem[{Maloney et~al(2010)Maloney, Resnick, Rusk, Silverman, and
  Eastmond}]{Maloney2010}
Maloney J, Resnick M, Rusk N, Silverman B, Eastmond E (2010) The scratch
  programming language and environment. ACM Transactions on Computing Education
  (TOCE) 10(4):16

\bibitem[{Mannila and de~Raadt(2006)}]{LindaMannilaMichaeldeRaadt2006}
Mannila L, de~Raadt M (2006) An objective comparison of languages for teaching
  introductory programming. In: Proceedings of the 6th Baltic Sea conference on
  Computing education research: Koli Calling 2006, ACM, pp 32--37

\bibitem[{Mannila et~al(2006)Mannila, Peltom{\"a}ki, and
  Salakoski}]{Mannila2006}
Mannila L, Peltom{\"a}ki M, Salakoski T (2006) What about a simple language?
  analyzing the difficulties in learning to program. Computer Science Education
  16(3):211--227

\bibitem[{Matsuzawa et~al(2015)Matsuzawa, Ohata, Sugiura, and
  Sakai}]{Matsuzawa2015}
Matsuzawa Y, Ohata T, Sugiura M, Sakai S (2015) Language migration in non-cs
  introductory programming through mutual language translation environment. In:
  Proceedings of the 46th ACM Technical Symposium on Computer Science
  Education, ACM, pp 185--190

\bibitem[{Meerbaum-Salant et~al(2011)Meerbaum-Salant, Armoni, and
  Ben-Ari}]{Meerbaum-Salant2011}
Meerbaum-Salant O, Armoni M, Ben-Ari M (2011) Habits of programming in scratch.
  In: Proceedings of the 16th annual joint conference on Innovation and
  technology in computer science education, ACM, pp 168--172

\bibitem[{Mendeley(2017)}]{Mendeley}
Mendeley (2017) {Mendeley Reference Manager}.
  \urlprefix\url{https://www.mendeley.com/}, [Online; accessed 17-March-2017]

\bibitem[{at~the MIT Media~Lab(2017)}]{ScratchStats}
at~the MIT Media~Lab LKG (2017) {Scratch Statistics}.
  \urlprefix\url{https://scratch.mit.edu/statistics/}, [Online; accessed
  09-March-2017]

\bibitem[{Montero et~al(2010)Montero, D{\'\i}az, D{\'\i}ez, and
  Aedo}]{Montero2010}
Montero S, D{\'\i}az P, D{\'\i}ez D, Aedo I (2010) Dual instructional support
  materials for introductory object-oriented programming: classes vs. objects.
  In: Education Engineering (EDUCON), 2010 IEEE, IEEE, pp 1929--1934

\bibitem[{Mullins et~al(2009)Mullins, Whitfield, and Conlon}]{Mullins2009}
Mullins P, Whitfield D, Conlon M (2009) Using alice 2.0 as a first language.
  Journal of Computing Sciences in Colleges 24(3):136--143

\bibitem[{Nikula et~al(2007)Nikula, Sajaniemi, Tedre, and Wray}]{Nikula2007}
Nikula U, Sajaniemi J, Tedre M, Wray S (2007) Python and roles of variables in
  introductory programming: experiences from three educational institutions.
  Journal of Information Technology Education 6:199--214

\bibitem[{Nolan and Bergin(2016)}]{nolan2016role}
Nolan K, Bergin S (2016) The role of anxiety when learning to program: a
  systematic review of the literature. In: Proceedings of the 16th Koli Calling
  International Conference on Computing Education Research, ACM, pp 61--70

\bibitem[{Parker(2011)}]{Parker}
Parker B (2011) Teaching experiences with alice for high school students.
  Journal of Computing Sciences in Colleges 27(2):148--155

\bibitem[{Price and Barnes(2015)}]{Price2015}
Price TW, Barnes T (2015) Comparing textual and block interfaces in a novice
  programming environment. In: Proceedings of the eleventh annual International
  Conference on International Computing Education Research, ACM, pp 91--99

\bibitem[{Quille et~al(2015)Quille, Bergin, and Mooney}]{Quille2015}
Quille K, Bergin S, Mooney A (2015) {Programming: Factors that Influence
  Success Revisited and Expanded}

\bibitem[{Ranade(2016)}]{Ranade2016}
Ranade AG (2016) Introductory programming: Let us cut through the clutter! In:
  Proceedings of the 2016 ACM Conference on Innovation and Technology in
  Computer Science Education, ACM, pp 278--283

\bibitem[{Robinson(2016)}]{Robinson2016}
Robinson W (2016) From scratch to patch: Easing the blocks-text transition. In:
  Proceedings of the 11th Workshop in Primary and Secondary Computing
  Education, ACM, pp 96--99

\bibitem[{Sandoval-Reyes et~al(2011)Sandoval-Reyes, Galicia-Galicia, and
  Gutierrez-Sanchez}]{Sandoval-Reyes2011}
Sandoval-Reyes S, Galicia-Galicia P, Gutierrez-Sanchez I (2011) Visual learning
  environments for computer programming. In: Electronics, Robotics and
  Automotive Mechanics Conference (CERMA), 2011 IEEE, IEEE, pp 439--444

\bibitem[{Seals et~al(2008)Seals, Mcmillian, Rouse, Agarwal, Johnson, Gilbert,
  and Chapman}]{Seals2008}
Seals C, Mcmillian Y, Rouse K, Agarwal R, Johnson AW, Gilbert JE, Chapman R
  (2008) Computer gaming at every age: A comparative evaluation of alice.
  i-Manager's Journal of Educational Technology 5(3):1

\bibitem[{da~Silva~Ribeiro et~al(2014)da~Silva~Ribeiro,
  de~Oliveira~Brand{\~a}o, Faria, and Brand{\"a}o}]{daprogramming}
da~Silva~Ribeiro R, de~Oliveira~Brand{\~a}o L, Faria TVM, Brand{\"a}o AAF
  (2014) Programming web-course analysis: how to introduce computer
  programming? In: Frontiers in Education Conference (FIE), 2014 IEEE, IEEE, pp
  1--8

\bibitem[{Slany(2012)}]{Slany2012}
Slany W (2012) Catroid: a mobile visual programming system for children. In:
  Proceedings of the 11th International Conference on Interaction Design and
  Children, ACM, pp 300--303

\bibitem[{Smith et~al(2014)Smith, Sutcliffe, and Sandvik}]{Smith2014}
Smith N, Sutcliffe C, Sandvik L (2014) Code club: bringing programming to uk
  primary schools through scratch. In: Proceedings of the 45th ACM technical
  symposium on Computer science education, ACM, pp 517--522

\bibitem[{Software(2017)}]{TIOBE}
Software T (2017) {TIOBE Index for March 2017}.
  \urlprefix\url{http://www.tiobe.com/tiobe-index/}, [Online; accessed
  15-March-2017]

\bibitem[{Sykes(2007)}]{Sykes2007}
Sykes ER (2007) Determining the effectiveness of the 3d alice programming
  environment at the computer science i level. Journal of Educational Computing
  Research 36(2):223--244

\bibitem[{Tangney et~al(2010)Tangney, Oldham, Conneely, Barrett, and
  Lawlor}]{Tangney2010}
Tangney B, Oldham E, Conneely C, Barrett S, Lawlor J (2010) Pedagogy and
  processes for a computer programming outreach workshop—the bridge to
  college model. IEEE Transactions on Education 53(1):53--60

\bibitem[{Weintrop(2015)}]{Sciences2015}
Weintrop D (2015) Blocks, text, and the space between: The role of
  representations in novice programming environments. In: Visual Languages and
  Human-Centric Computing (VL/HCC), 2015 IEEE Symposium on, IEEE, pp 301--302

\bibitem[{Weintrop and Wilensky(2015)}]{Weintrop2015}
Weintrop D, Wilensky U (2015) To block or not to block, that is the question:
  students' perceptions of blocks-based programming. In: Proceedings of the
  14th International Conference on Interaction Design and Children, ACM, pp
  199--208

\end{thebibliography}

\end{document}